\documentclass[aip, jmp, amsmath, amssymb, reprint]{revtex4-1}

\usepackage{graphicx}
\usepackage{bm}
\usepackage[german,english]{babel}
\usepackage{natmove}

\begin{document}

\title{Understanding three-body contributions to coarse-grained force-fields}

\author{Christoph Scherer}
\email{scherer@mpip-mainz.mpg.de}
\affiliation{Max Planck Institute for Polymer Research, Ackermannweg 10, 55128 Mainz, Germany}

\author{Denis Andrienko}
\email{denis.andrienko@mpip-mainz.mpg.de}
\affiliation{Max Planck Institute for Polymer Research, Ackermannweg 10, 55128 Mainz, Germany}

\date{\today}

\begin{abstract}
Coarse-graining (CG) is a systematic reduction of the number of degrees of freedom (DOF) used to describe a system of interest. CG can be thought of as a projection on coarse-grained DOF and is therefore dependent on the functions used to represent the CG force field. In this work, we show that naive extensions of the coarse-grained force-field can result in unphysical parametrizations of the CG potential energy surface (PES). This issue can be elevated by coarse-graining the two- and three-body forces separately, which also helps to evaluate the importance of many-body interactions for a given system. The approach is illustrated on liquid water, where three-body interactions are essential to reproduce the structural properties, and liquid methanol, where two-body interactions are sufficient to reproduce the main features of the atomistic system.
\end{abstract}

\maketitle

Coarse-graining (CG) is a systematic way of reducing the number of degrees of freedom describing a specific physical system. A typical but by no means complete list of coarse-graining procedures includes (i) a renormalization group analysis in the vicinity of a critical point, where degrees of freedom (e.g. spins) are blocked together~\cite{zinn-justin_quantum_2002}; (ii) the formulation of system dynamics in terms of a master equation, with the entire phase space represented by a few states~\cite{jansen_monte_1995}; (iii) parametrizations of classical force-fields, in which electronic degrees of freedom are incorporated into classical interaction potentials~\cite{jorgensen_opls_1988}; (iv) united-atom or coarser classical particle-based force fields, with light atoms (e.g. hydrogens) incorporated into the heavier ones~\cite{moral_cost-effective_2015}. 

Coarse-graining often relies on a certain time-scale separation: due to some parts of the system evolving on a significantly slower timescale than others. For example, diffusion and rotation of molecules, or parts of molecules, is a much slower process than a characteristic bond vibration. It is then possible to combine several coherently moving atoms into a single interaction site. This reduces the computational costs in several ways. First, the coarse-grained system has less degrees of freedom. Second, smoother (softer) interaction potentials result in a smaller friction, hence, faster dynamics, which can now be propagated using a bigger simulation time step. Though the connection between atomistic and coarse-grained dynamics is rather non-trivial~\cite{izvekov_mori-zwanzig_2017, lange_collective_2006, hijon_morizwanzig_2009, rudzinski_communication:_2016}, it often helps to reach ten to a hundred times longer simulation times. 

The coarse-graining procedure in itself involves three steps: choice of coarse-grained degrees of freedom and a mapping of the fine- to the coarse-grained description, identification of a merit function which quantifies the difference between the fine- and coarse-grained representations, and determination of the coarse-grained PES. Consistency between the coarse-grained and the fine-grained models requires consistency of the equilibrium probability densities which, given a specific mapping, results in unique expressions for the CG masses and interaction potential~\cite{noid_multiscale_2008}. Evaluating this many-body potential of mean force (PMF) is, however, as computationally demanding as propagating the fine-grained system, counteracting the idea of coarse-graining to reduce computational cost. In practice, the many-body PMF is approximated with or, in other words, projected onto basis functions that are used to represent the coarse-grained bonded (bond, angle and dihedral) and non-bonded interactions. 

By choosing an appropriate projection operator one can, for example, reproduce structural quantities of the atomistic system. Iterative Boltzmann inversion (IBI)~\cite{reith_deriving_2003}, inverse Monte Carlo (IMC)~\cite{lyubartsev_reverse_1995}, or relative entropy minimization~\cite{shell_relative_2008} schemes have been developed for this purpose. An alternative route is to match the forces of the CG system to the ones of the atomistic description, employing the force matching (FM) or multiscale coarse-graining (MS-CG) methods~\cite{izvekov_effective_2004, noid_multiscale_2008}. FM can also be connected to structure based CG via YBG theory~\cite{mullinax_generalized-yvonborngreen_2010, rudzinski_generalized-yvon-born-green_2015}. 

Non-bonded interaction potentials of practically all classical force-fields are represented by pair potentials. This limits the CG force-field functions in the out-of the box molecular dynamics packages, and hence the accuracy of the CG model. Indeed, for locally structured liquids, where liquid water is a prominent example, as well as mixtures, CG pair potentials fail to capture the structural properties~\cite{larini_multiscale_2010, das_multiscale_2012, das_multiscale_2009, jochum_structure-based_2012, scherer_comparison_2016}. In this respect the many-body terms become important: by adding a short-ranged non-bonded three-body potential one can significantly improve the structural and thermodynamic properties~\cite{molinero_water_2009, larini_multiscale_2010, das_multiscale_2012, zipoli_improved_2013, lu_coarse-graining_2014, cisneros_modeling_2016, das_multiscale_2012-1}. An alternative way of incorporating many-body effects is by using local density dependent potentials~\cite{sanyal_coarse-grained_2016} 
or by using machine learning to predict many-body 
contributions~\cite{bartok_gaussian_2015}.

\begin{figure*}[htbp]
\includegraphics[scale=1.0]{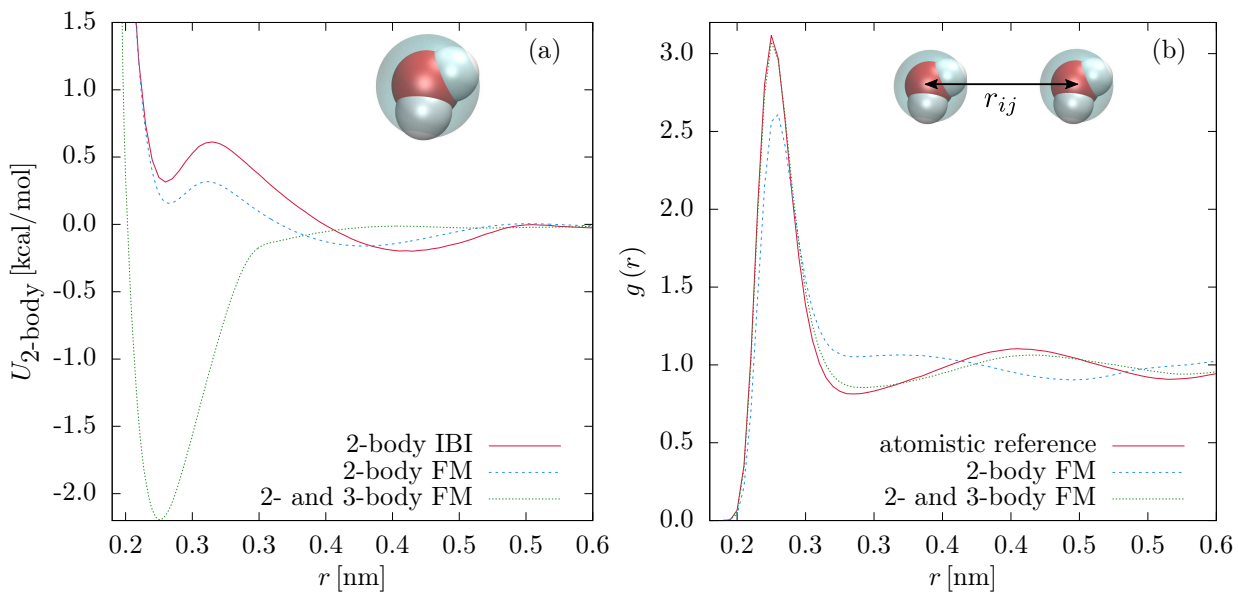}
\caption{(a) the pair potentials and (b) the radial distribution functions of coarse-grained SPC/E water. Three parametrization are shown: structural matching (IBI), force matching using (tabulated) pair potentials only, and force-matching with tabulated pair potentials and the short-range three-body SW potential. The CG IBI radial distribution function perfectly matches the atomistic reference curve by construction.}
\label{fig:Fig1}
\end{figure*}

Let us, however, re-examine the effect of short-ranged three-body potentials, such as the three-body Stillinger-Weber (SW) potential~\cite{stillinger_computer_1985},
\begin{align}
\nonumber
  U = \sum_{i,j\neq i,k>j}\,f \left(\theta_{ijk}\right) \exp{\left(\frac{\gamma_{ij}\sigma_{ij}}{r_{ij}-a_{ij}\sigma_{ij}} + \frac{\gamma_{ik}\sigma_{ik}}{r_{ik}-a_{ik}\sigma_{ik}}\right)}, 
  \label{eq:SW1}
\end{align}
on the structure and thermodynamics of liquid water~\cite{molinero_water_2009,larini_multiscale_2010,zipoli_improved_2013,lu_coarse-graining_2014, cisneros_modeling_2016}. Note that in our case $f \left(\theta_{ijk}\right)$ is a tabulated function of $\theta_{ijk}$.
In what follows, all coarse-grained MD simulations are performed using the LAMMPS package~\cite{plimton_fast_1995}, atomistic simulations using the GROMACS~\cite{abraham_gromacs:_2015} package, and coarse-graining using the VOTCA package~\cite{ruhle_versatile_2009}. We employ the SPC/E water model~\cite{berendsen_missing_1987} and the OPLS-AA force field~\cite{jorgensen_opls_1988,jorgensen_development_1996} without constraints for methanol. More details are provided in the Supporting Information.

Fig.~\ref{fig:Fig1}(b) illustrates how the incorporation of the three-body potential leads to a significant improvement of the CG model compared to the CG model with only two-body FM interactions. In fact, not only the structure, represented by the radial distribution function in Fig.~\ref{fig:Fig1}(b), but also the thermodynamic properties are significantly improved after extending the CG basis set. It turns out, however, that the addition of the SW potential leads to a significantly more attractive two-body potential (see Fig.~\ref{fig:Fig1}(a), dotted green line) as compared to the CG models with pair potentials only (see Fig.~\ref{fig:Fig1}(a), solid red line (IBI) and dashed blue line (FM))~\cite{ruhle_versatile_2009, larini_multiscale_2010}. The same holds for liquid methanol which is shown in the supporting information. This is surprising, since the corresponding peaks in the radial distribution functions have the same height (solid red and dotted green lines in Fig.~\ref{fig:Fig1}(b)). One might even 
jump into conclusion that the three-body term is as important as the two-body one, i.e. it cannot be treated as a part of a perturbative expansion of PMF into the many-body potentials. 

In order to understand this discrepancy, let us examine the pair potential of mean force, $U_{\text{PMF}}\left(r\right)= - \int_0^r \, F_r\left(r^\prime\right)\,\text{d}r^\prime$ between two coarse-grained sites. Here $F_r\left(r\right)$ is the radial component of the total force on a CG bead averaged over all pairs of CG beads with distance $r$. This force is evaluated in the CG simulation run for different CG interactions or in the atomistic simulation by employing the CG mapping. Note that $U_\text{PMF}(r)$ is equivalent to the radial distribution function, since $g(r) \sim \exp(-U_\text{PMF}(r)/k_\text{B}T)$. It is, however, easier to interpret its decomposition on two- and three-body contributions. 

\begin{figure*}
\includegraphics[scale=1.0]{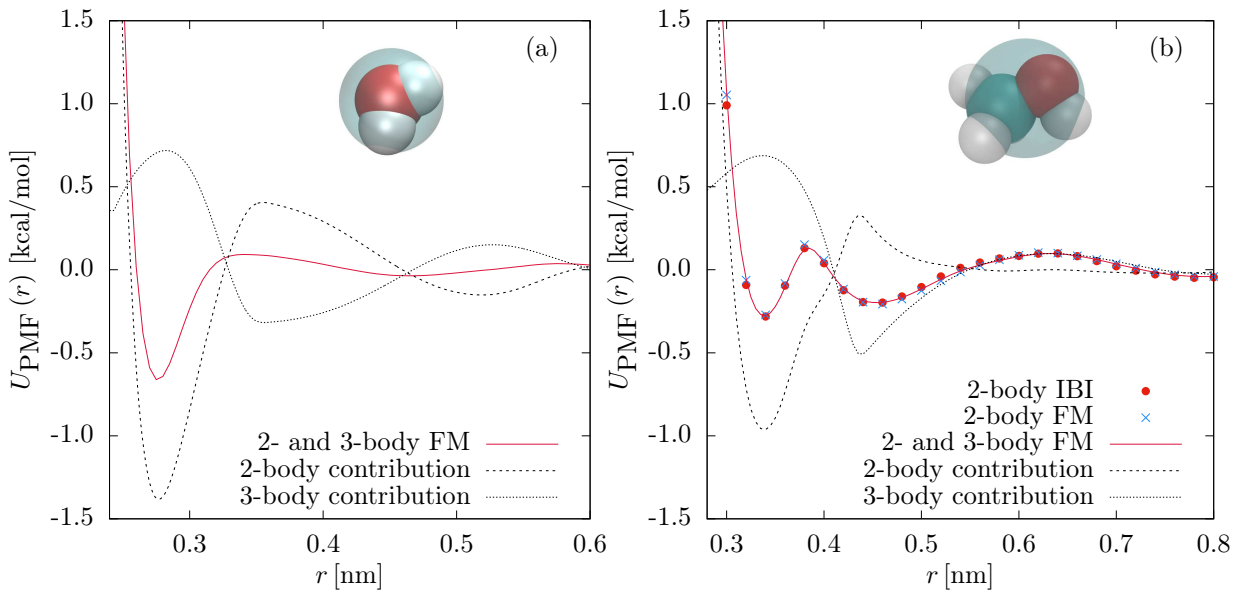}
\caption{Splitting of CG two-body potentials of mean force for (a) SPC/E water and (b) liquid methanol. 2-body IBI, 2-body force matching, 2- and 3-body force matching refer to the three different CG parametrizations: the iterative Boltzmann inversion of the radial distribution function, force-matching with a tabulated pair potential, and force matching with the tabulated pair potential, as well as, the three-body SW potential.}
\label{fig:Fig2}
\end{figure*}

Such a decomposition is shown in Fig.\ref{fig:Fig2} for liquid water and methanol. Here, strong attractive forces manifest themselves in a large dip in the two-body contribution to the PMF within the first coordination shell (dashed line). This attraction is, however, compensated by a short-range repulsive force coming from the three-body SW interaction (dotted line). Similar compensation can be observed for liquid methanol. Here, both the radial distribution function and the two-body PMF are perfectly reproduced by the coarse-gained force-fields with or without the inclusion of the three-body term (parametrizations using pair potentials only are shown as crosses and dots in Fig.~\ref{fig:Fig2}(b)). Hence, one might anticipate that the contribution of the three-body term should be very small, which is obviously not the case. This indicates that the SW three-body potential is ``not orthogonal'' to the pair potential when the liquid structure is chosen as a target observable.

This results in the motivation to test a different parametrization scheme. Following the idea of the Gram-Schmidt orthogonalization, we now parametrize the three-body SW potential using the residual force $\Delta \bm{f}_i$, i.e., after subtracting the two-body force from the total force on each CG bead $i$: $\Delta \bm{f}_i=\bm{f}_i^{\text{ref}}-\bm{f}_i^{\text{2-body}}$. The two-body force is the force obtained using either the FM or the IBI parametrization with pair potentials only.

\begin{figure*}
\includegraphics[scale=1.0]{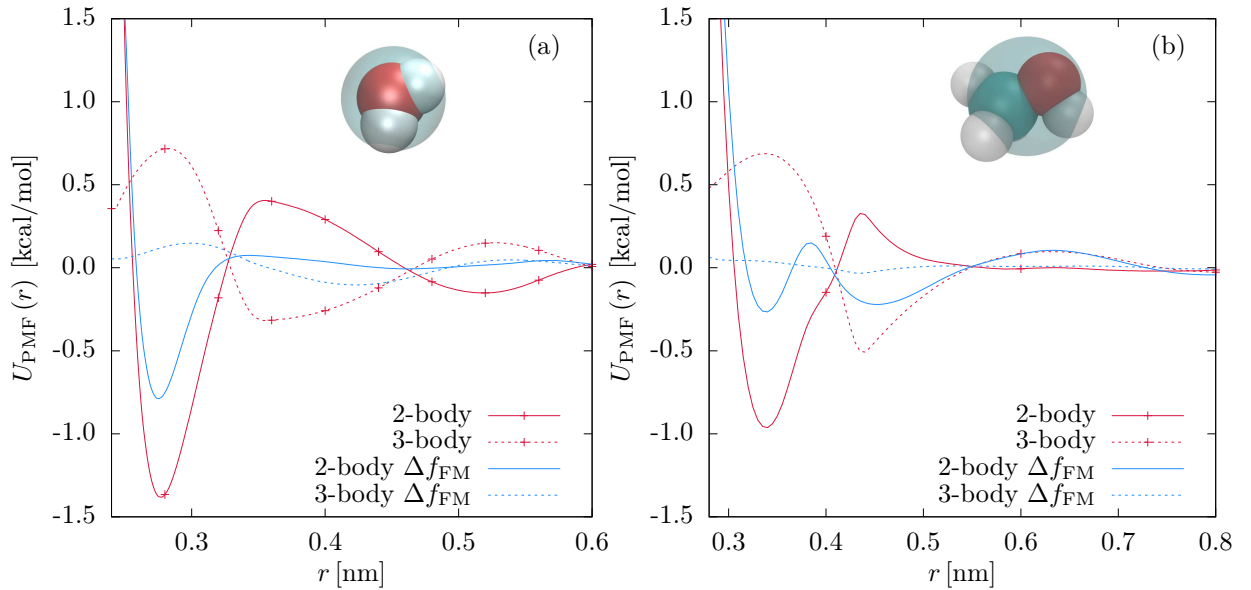}
\caption{Splitting of the CG two-body potentials of mean force for different CG parametrization schemes for (a) SPC/E water and (b) methanol. The results shown are for the concurrent two-body and three-body FM parametrization and the three-body FM parametrization using the residual force, $\Delta f_\text{FM}$. The same trend is observed for the IBI pair force,  $\Delta f_\text{IBI}$. }
\label{fig:Fig3}
\end{figure*}

\begin{figure*}
\includegraphics[scale=1.0]{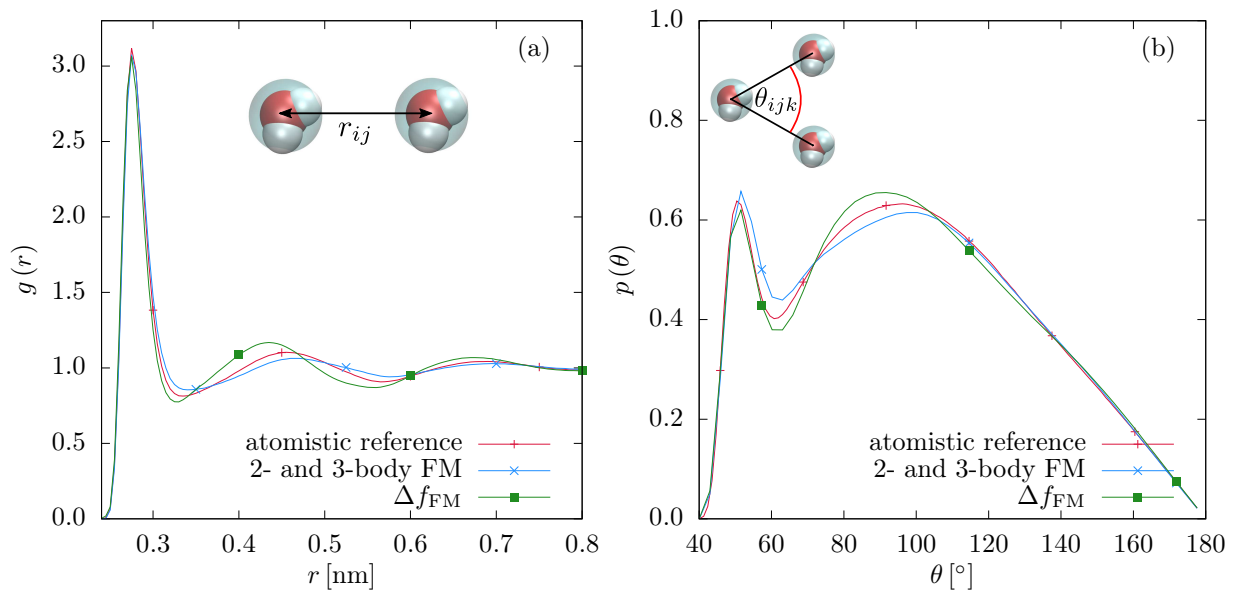}
\caption{(a) Radial and (b) angular distribution functions for different CG potentials of SPC/E water. The angular distribution function was calculated for the first coordination shell using a cutoff of $0.37\,\text{nm}$. }
\label{fig:Fig4}
\end{figure*}

The results are shown Fig.~\ref{fig:Fig3}. Both for water and methanol, the parametrization of the residual force significantly reduces the three-body contribution to the two-body PMF.  As a result, the attractive part of the two-body contribution to PMF is also reduced. It is of course important to check whether the parametrization using the residual force is still capable of reproducing the liquid structure. In Fig.~\ref{fig:Fig4}, we compare the radial distribution functions and the angular distributions of CG beads within the first coordination shell for several parametrization schemes for liquid water. The completely unconstrained parametrization performs slightly better than the one with the residual force, but the overall agreement (especially taking into account the results without any three-body contributions) is satisfactory. In fact, for liquid methanol (see the supporting information) the agreement is very good. 

\begin{table*}
\begin{tabular}{ |c|c|c|c|c|c|c|c|c| }
\hline
 & & 2-body IBI & 2-body FM & 2-and 3-body FM & $\quad\Delta f_\text{IBI}\quad$ & $\quad\Delta f_\text{FM}\quad$ & all-atom & experiment \tabularnewline
\hline 
  water & $p$ [kbar] & 7.56 & 7.18 & 0.87 & 6.46 & 4.16 & 0.001 & 0.001 \tabularnewline
\cline{2-9}
    & $\Delta H$ [kcal/mol] & -2.74 & -2.52 & 4.01 & -1.58 & -0.39 & 11.76 & 10.52 \tabularnewline
\hline 
  methanol & $p$ [kbar] & 4.17 & 3.61 & 0.82 & 4.53 & 3.92 & 0.001 & 0.001 \tabularnewline
\cline{2-9}
   & $\Delta H$ [kcal/mol] & -5.30 & -4.06 & 2.56 & -5.67 & -4.37 & 8.94 & 8.95 \tabularnewline
\hline
\end{tabular}
\caption{Comparison of enthalpy of vaporization $\Delta H$ and internal pressure $p$ of the different CG models for NVT simulations at the atomistic density. Experimental values of $\Delta H$ for water and methanol can be found in Refs.~\onlinecite{lu_coarse-graining_2014, caleman_force_2012}.}
\label{tab:thermodynamics}
\end{table*}

It is known that the incorporation of three-body terms improves the thermodynamic properties of CG water~\cite{molinero_water_2009,larini_multiscale_2010,lu_coarse-graining_2014}. Here, we assess the performance of the CG models in terms of the internal pressure $p$ and the enthalpy of vaporization $\Delta H$. Details can be found in the supporting information. The internal (virial) pressures of all CG models are given in Table~\ref{tab:thermodynamics}. Ideally, the pressure of the CG models (simulated at the density of the atomistic model) should be $1\,\text{atm}$. It can be clearly seen that the CG models with two- and three-body interactions parametrized simultaneously still yield the best results, both for water and methanol. The pressure reduction, to a large extent, comes from the significantly more attractive pair potential (see Fig.~\ref{fig:Fig1}). In fact, for liquid methanol the addition of the short-ranged three-body potential parametrized using the residual force even leads to slightly inferior results compared to both two-body parametrizations. Note that we do not use any pressure corrections nor any explicit pressure matching schemes~\cite{dunn_bottom-up_2015}.

We also evaluate the enthalpy of vaporization, for which experimental values are readily available~\cite{lu_coarse-graining_2014, caleman_force_2012}. Table~\ref{tab:thermodynamics} lists the values of the enthalpy of vaporization taking into account the actual average liquid pressure of each CG parameterization. Essentially, the same conclusions can be drawn as comparing the pressure of the different CG models: The best values are obtained for the CG models with two- and three-body interactions parametrized simultaneously. Note that the overall discrepancies of the enthalpies of vaporization can not be attributed to the atomistic force fields (SPC/E and OPLS) as the all-atom results are close to the experimental values in accordance with previous work~\cite{caleman_force_2012,lu_coarse-graining_2014}.

To conclude, we have shown that adding the Stillinger-Weber three-body potential to the coarse-grained force-field leads to a strongly attractive pair interaction potential. Short-range attraction is then compensated by the three-body contribution in a way that the total pair potential of mean force does not change. Parametrization of the three-body term with residual forces reduces the three-body contribution to the pair potential but at the same time worsens thermodynamic properties of the coarse-grained model, quantified by pressure and vaporization enthalpy. Our work indicates that the three-body Stillinger-Weber potential is not optimal for capturing three-body interactions in soft condensed matter systems. The proposed approach helps to better understand the importance of many-body terms in the coarse-grained force field and paves the way to the development of computationally efficient many-body potentials tailored to reproduce both thermodynamic and structural properties.

\acknowledgments
We thank Joseph Rudzinski, Tristan Bereau, Leanne Paterson, and Kurt Kremer for fruitful discussions and revision of the manuscript. DFG is acknowledged for financial support through the collaborative research center TRR 146.

\bibliographystyle{apsrev4-1}
\bibliography{literature_short}

\end{document}